\documentclass[11pt]{article}
\usepackage{amsmath}
\usepackage{graphicx}% Include figure files
\usepackage{bm}% bold math
\usepackage{pstricks}
\newcommand{\beq}{\begin{eqnarray*}}
\newcommand{\eeq}{\end{eqnarray*}}

\newcommand{\bp}{\begin{minipage}[c]{4ex}\begin{tabular}{c}}
\newcommand{\ep}{\end{tabular}\end{minipage}}

\newcommand{\be}{\begin{equation}}
\newcommand{\ee}{\end{equation}}

\begin{document}

% Title of the article
\title{Mechanisms of   doping  graphene}

% Authors
\author{ 
  H.~Pinto$^1$, R.~Jones$^1$, J.~P. Goss$^2$, P.~R.~Briddon$^2$ \\
  \\\vspace{6pt} 
  $^1$School of Engineering, Mathematics and Physical Sciences,\\ University of Exeter, EX4~4QL, UK\\
  $^2$School of Electrical, Electronic and Computer Engineering,\\ Newcastle University, Newcastle upon Tyne, NE1~7RU, UK\\
}

\maketitle

\abstract{
We distinguish  three mechanisms of doping graphene.
Density functional theory is used to show 
that  electronegative  molecule like F4-TCNQ  and
electropositive  metals like K  dope graphene p- and n-type
respectively.
These dopants are expected to lead to a decrease in carrier mobility arising from
Coulomb scattering  but  without any hysteresis effects.
Secondly, a novel doping mechanism is exhibited by Au which dopes
bilayer graphene but not single layer. Thirdly, electrochemical doping
is effected by redox reactions and can result in p-doping by
humid atmospheres and n-doping by NH$_3$ and toluene.
}

\section{Introduction}
The type and concentration of carriers in graphene can be controlled by
adsorbates or substrates which transfer charge to graphene. 
$p$-type electronic doping results from a disparity between the 
work function  of graphene and the electron affinity of the adsorbate.
In the simplest case this is caused by  a  difference  between the 
highest  occupied molecular level (HOMO)  and the lowest unoccupied molecular level (LUMO)  of the adsorbate and graphene. This alignment
as we shall see can  depend on the substrate as we shall show that 
Au dopes  bilayer graphene but is ineffective for single layer 
graphene.     Electrochemical doping results from redox
reactions  near the graphene surface. The timescales for the two types of doping is different since electronic
doping occurs spontaneously while electrochemical doping requires longer   times to overcome reaction and diffusion barriers.
Hysteresis effects for graphene based  field effect devices   can  also be  expected  in electrochemical doping since the concentration of the reactants  
can change at a rate slower  than the change in gate voltages.  

Electropositive elements like metals are good candidates for electronic n-type doping of graphene. 
Molecular beam deposition of transition metal clusters, Ti, Fe and Pt, 
has shown that these metals effect n-doping of  graphene
whereas bulk deposition of Pt dopes  graphene p-type\cite{pi}. Electronic doping on graphene was also 
demonstrated with K atoms deposited at low temperatures (20K) on graphene in ultrahigh vacuum (UHV)\cite{j.h.chen}. 
Synchrotron-based high resolution photoemission spectroscopy (PES) showed  that graphene can be made p-type by the deposition of tetrafluoro-tetracyanoquinodimethane (F4-TCNQ).
Because of the molecular separation of the adsobate and graphene,
all these dopants are expected to lead to a decrease in carrier mobility arising from
Coulomb scattering without any hysteresis effects.

Exposure of graphene  to a humid atmosphere  or gaseous nitrogen dioxide leads to p-type behaviour,
while ammonia leads to  n-type behaviour
\cite{schedin-NM-6-652}. A great surprise is  that  it was found, in particular experimental conditions, gaseous
toluene leads to n-doping of graphene\cite{Kaverzin}, in contradiction with theoretical calculations which predict no transfer of
charge between graphene and the toluene molecule \cite{pinto}.
In this paper density functional calculations are used   to investigate the transfer of charge between graphene and adsorbates.
A electrochemical model for doping of graphene is also presented.

\section{Method}
The electronic ground state of graphene with different adsorbates was calculated using 
spin-optimised  density functional theory as implemented in the AIMPRO code\cite{aimpro,aimpro2}.
The exchange correlation potential was described using the local density approximation (LDA). 
The core electron were treated using the Hartwigsen Goedecker Hutter (HGH) pseudopotencials\cite{hgh}. 
The orbitals of the valence electrons consist of independent $s-, p-, d-$ like Gaussian functions centred
on atoms\cite{basis}. The electronic levels were filled using Fermi-Dirac statistics with $k_BT$=0.01~eV 
and a  metallic filling.
The Brillouin zone was sampled with a grid of 8$\times$8$\times$1 {\textbf k}-points within the 
Monkhorst-Pack scheme\cite{mpack}. 
Charge densities were Fourier-transformed using plane waves with an energy cut-off of 200 Ha.
The spin-populations of the cell were optimised by starting from random spin distributions. 
We used two different graphene unit cells, 4$\times$4$\times$1  and 6$\times$6$\times$1, 
enclosing 32 and 72 carbon atoms, respectively. 
During the relaxation all the atoms were allowed to move to their equilibrium positions.
The influence of the cell size was also studied with larger supercells and the band structure obtained and 
the consequent charge transfer  were very similar.

\section{Electronic doping}
The geometry of the relaxed   F4-TCNQ molecule was  found to be  in good agreement with 
experimental results  for  its precursor TCNQ \cite{pinto-JPCM-36-364220}. 
The electron affinity, the difference between the LUMO  and the vacuum level,
 was found to be 5.25 eV  in  agreement with an  experimental measurement
of  5.2 eV\cite{khan}. This deep LUMO level when compared
with the work function of graphene $\sim 4.5$ suggest that F4-TCNQ is a good candidate to act as an electronic acceptor on graphene.

The F4-TCNQ molecule was placed on top of a 72 carbon atoms graphene supercell, 6$\times$6$\times$1, and during the 
relaxation all the atoms were allowed to move. In the minimum energy configuration the molecule remain parallel to the 
graphene layer with a interplanar  separation  of 3.1 \AA. The binding energy was found to be 1.26 eV  which is probably  overestimated as a result  
of  the  LDA approximation.

\begin{figure}
\begin{center}
a)\includegraphics[angle=0,width=0.9\columnwidth,clip=true]{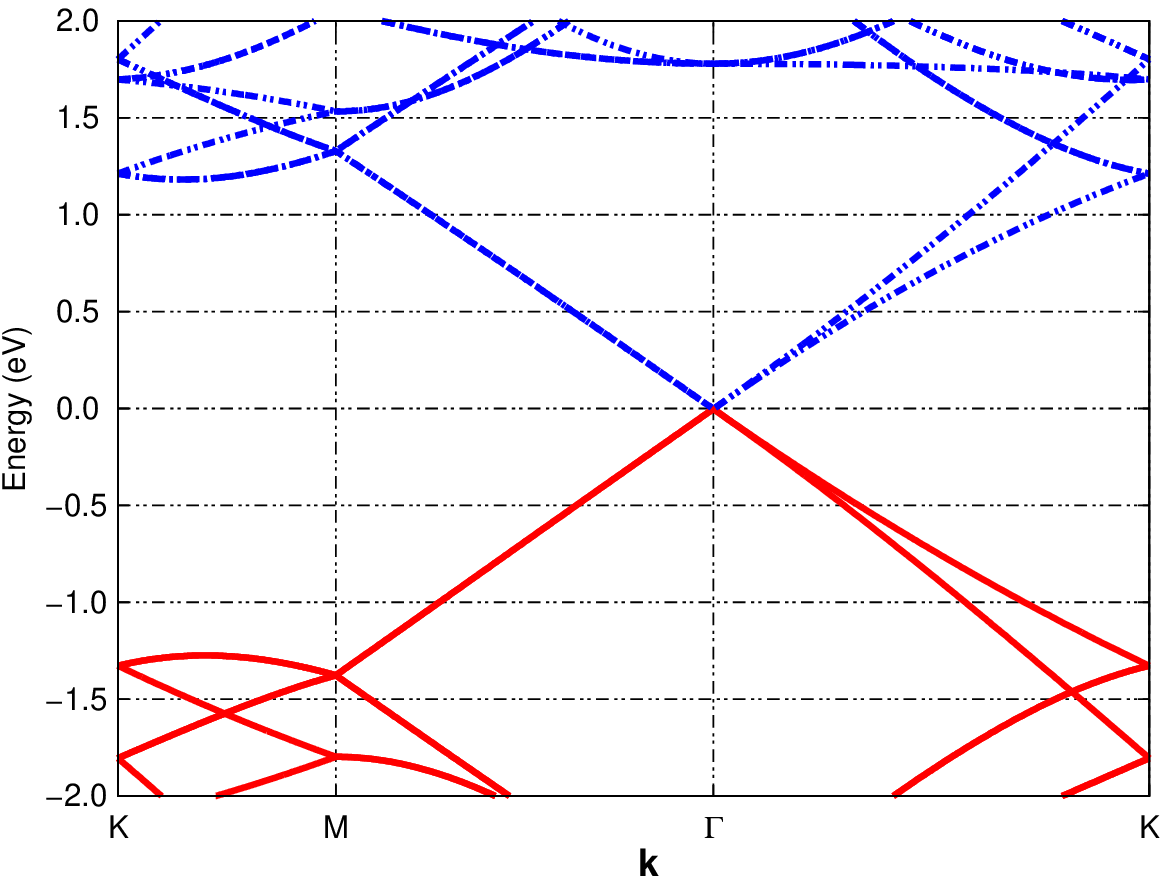}
b)\includegraphics[angle=0,width=0.9\columnwidth,clip=true]{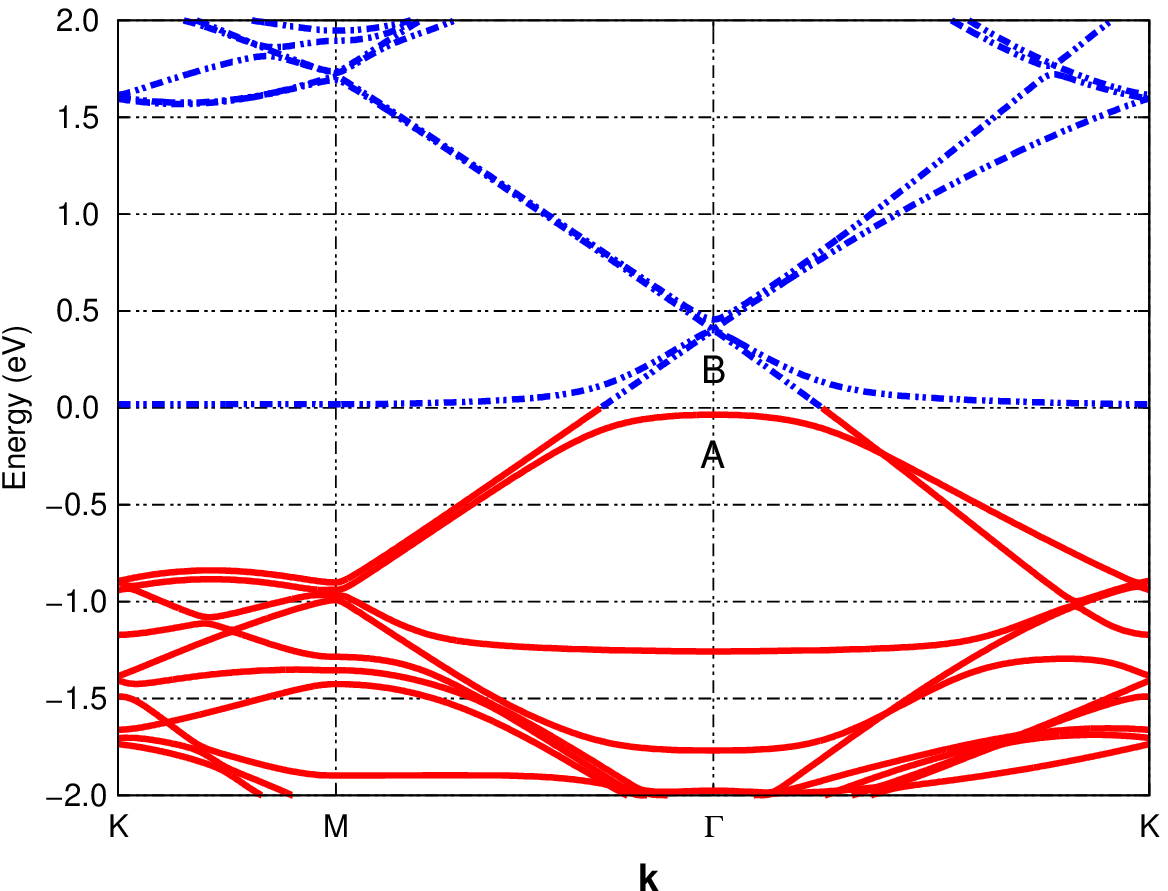}
\caption{\label{bandstructure}(Colour online)
Band structure (eV) of a) pristine graphene and b) F4-TCNQ on top of graphene plotted in the vicinity
of the Fermi energy along the high symmetry branches of the graphene Brillouin-zone. The Fermi level is
set to zero. Full lines (red) denote occupied states while dashed lines (blie) show empty levels. The Fermi level
is placed at zero. The curves shows the unoccupied F4-TCNQ levels around 0 eV become occupied near the
Dirac point indicating charge transfer. Colour online}
\end{center}
\end{figure}

\begin{figure}
\begin{center}
a)\includegraphics[angle=0,width=0.9\columnwidth,clip=true]{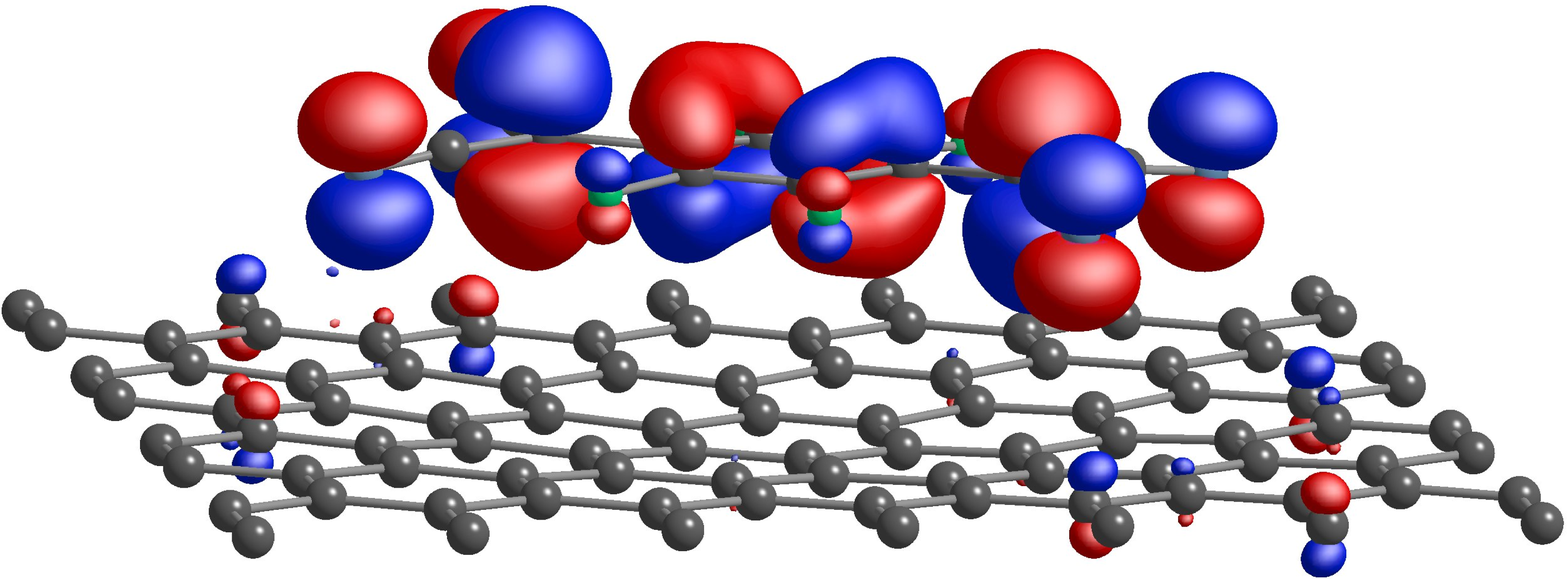}
b)\includegraphics[angle=0,width=0.9\columnwidth,clip=true]{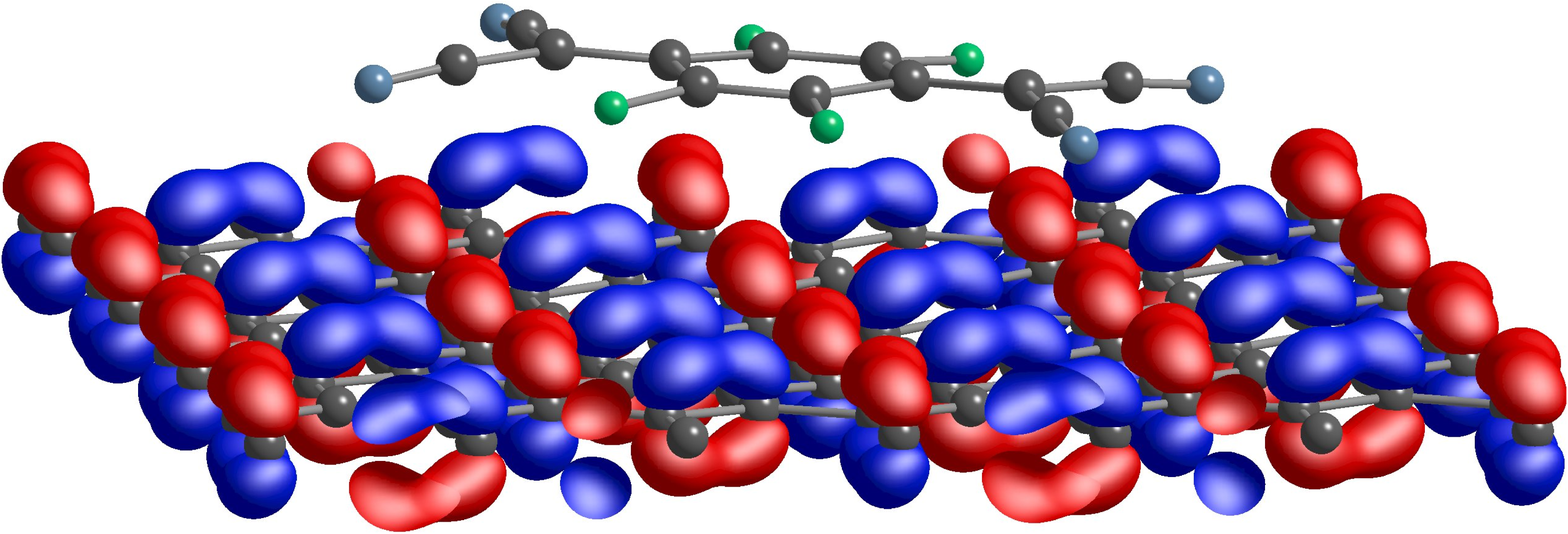}
\caption{\label{wavefunction}(Colour online)
Plot of the wavefunction of the a) HOMO level at $\Gamma$ marked A in figure \ref{wavefunction}b) shows strong
localisation on F4-TCNQ and
the b) LUMO at $\Gamma$ marked as B in figure \ref{wavefunction}a) shows
 strong  delocalisation  of a pi-bonding orbital over graphene but avoids F4-TCNQ.
Red and blue lobes are of equal amplitude and opposite sign.}
\end{center}
\end{figure}

Figure \ref{bandstructure} a) and b) shows respectively the electronic band structures of pristine graphene and graphene
with an adsorbed  molecule.
Because of band folding, the Dirac point, where the $\pi$ and $\pi^{\ast}$  bonding and anti-bonding  levels 
of  graphene meet, occurs at $\Gamma$ instead of the K point in the 6$\times$6$\times$1 
graphene supercell used here.    
Comparison  of  the two electronic band structures shows that when the molecule is placed on top of graphene,
%, Figure \ref{bandstructure} b),
the Fermi level is shifted to lie  {\it {below}} the Dirac point, and a new filled  flat band appears which is related to the adsorbate.  
This suggests transfer of negative
charge from graphene to F4-TCNQ
This  is supported by  an  analysis of the wavefunction 
shown in Figure \ref{wavefunction} a) of the level  marked A  in  Figure \ref{bandstructure} b). This  shows  that the  new filled band is strongly 
localised on the F4-TCNQ molecule. In addition,  the empty level marked  B,
which was occupied in graphene,  is delocalised over the graphene layer. 
This confirms that charge transfer between graphene and F4-TCNQ occurs.
The position of the Fermi-level indicates about  0.3 electrons are 
transferred  from graphene to a molecule of F4-TCNQ.   

To study the doping properties of metals, 
a single atom of Au and K respectiverly  were  placed on graphene in the three  obvious adsorption sites: above the centre 
of the graphene hexagonal ring, above a C-C bond and above the carbon atom. In the case of Au the minimum energy configuration
was found to be 2.27 \AA~on top of a carbon atom whereas K sits 2.42 \AA~directly above the middle of a hexagon optimising the number
of K-C bonds. Thus the K atom is strongly bound to graphene with a binding energy estimated to be 1.51 eV in contrast with the 0.65 eV
estimated for a  Au atom. This is in part due to their different doping properties. 
Spin-optimization led to a spin-less ground state for K and a
spin-1/2 state for Au. 

\begin{figure}
\begin{center}
 \includegraphics[width=0.9\columnwidth,clip=true]{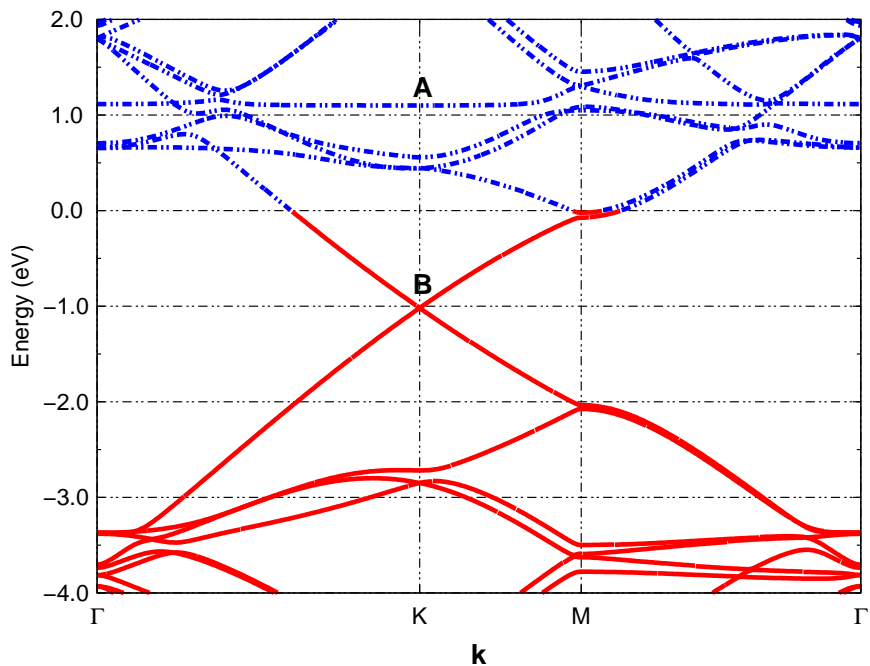}
\caption{\label{K_bandst}(Colour online)
Spin-averaged  electronic band structure (eV) of K on top of graphene in the vicinity of the Fermi energy.
The Fermi level is set to zero. Full lines denote occupied states while dashed lines show empty levels.
The bands around B, unoccupied for pristine graphene,  are now occupied.
}
\end{center}
\end{figure}

\begin{figure}
\begin{center}
a) \includegraphics[width=0.9\columnwidth,clip=true]{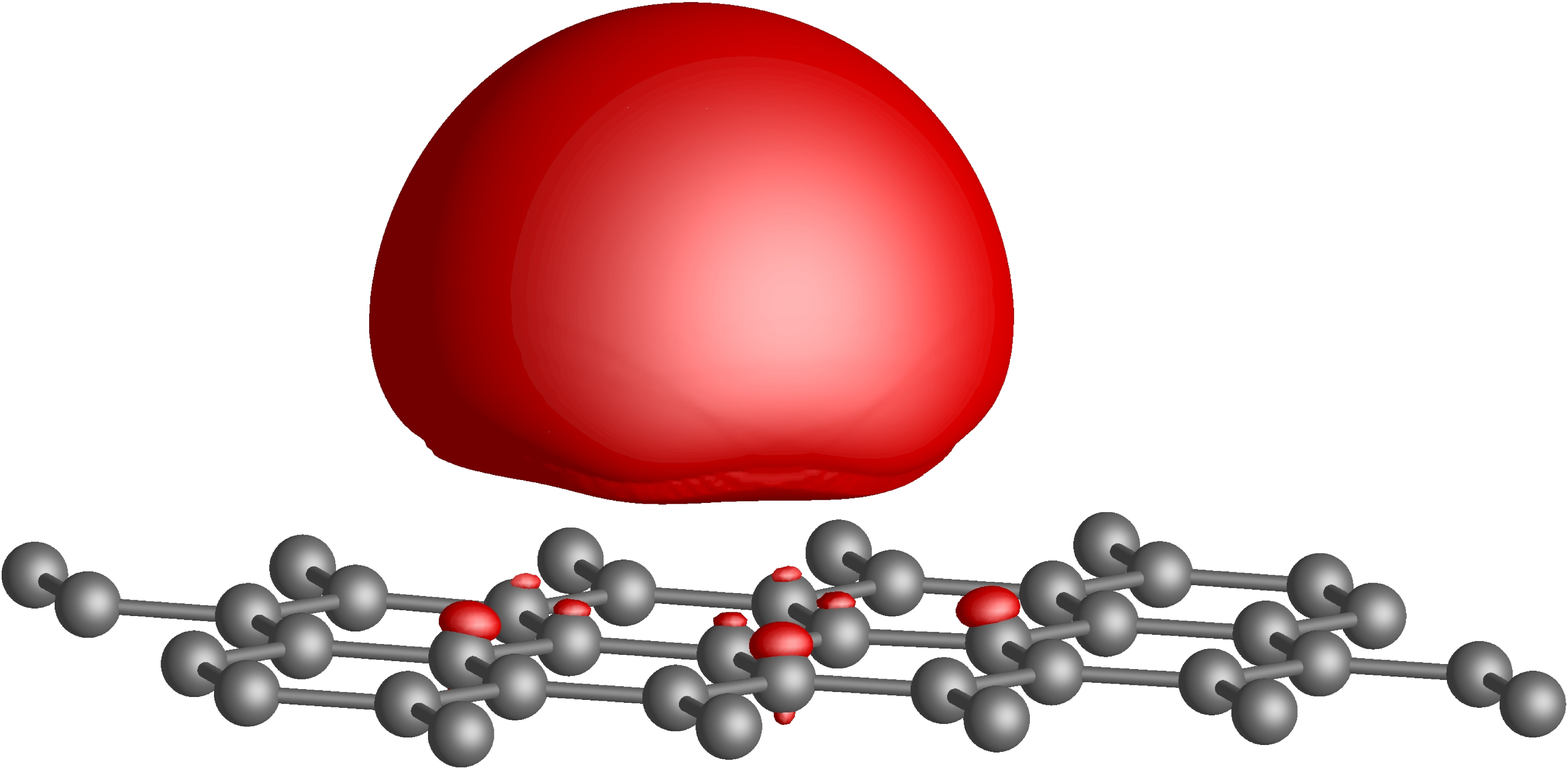}\\
b) \includegraphics[width=0.9\columnwidth,clip=true]{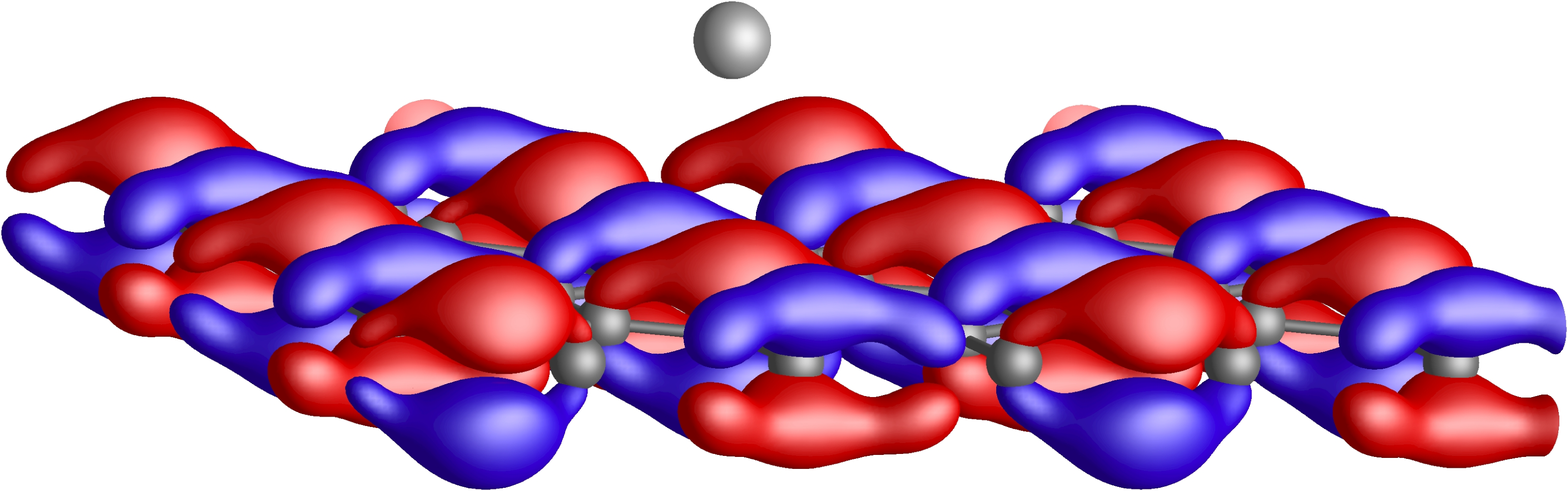}
\caption{\label{K_wave}(Colour online)
Plot of the wavefunctions of the majority spin electronic levels of a K atom  on top of graphene
at points marked a) A and b) B in figure \ref{K_bandst}. The wavefunction of the level marked A
is localised on the K atom while the wavefunction of the  highest  occupied level, marked B,
is delocalised over the graphene layer.
}
\end{center}
\end{figure}

Figure \ref{K_bandst} shows the electronic band structure of K on graphene. The Fermi level lies above the Dirac point which  indicates
doping of graphene with electrons. The wavefunction, shown in Figure \ref{K_wave} a), of the empty flat level marked A in Figure \ref{K_bandst} is strongly
localised on the K atom and  is derived from 
the 4$s$-level of K.  The wavefunction (Figure \ref{K_wave} b)) of the  filled level above the Dirac point, 
marked  B is  delocalised over the graphene layer.
This confirms that K  dopes  graphene $n$-type and we estimate about one electron per K atom is
transferred.

\begin{figure}
\begin{center}
a) \includegraphics[width=0.9\columnwidth,clip=true]{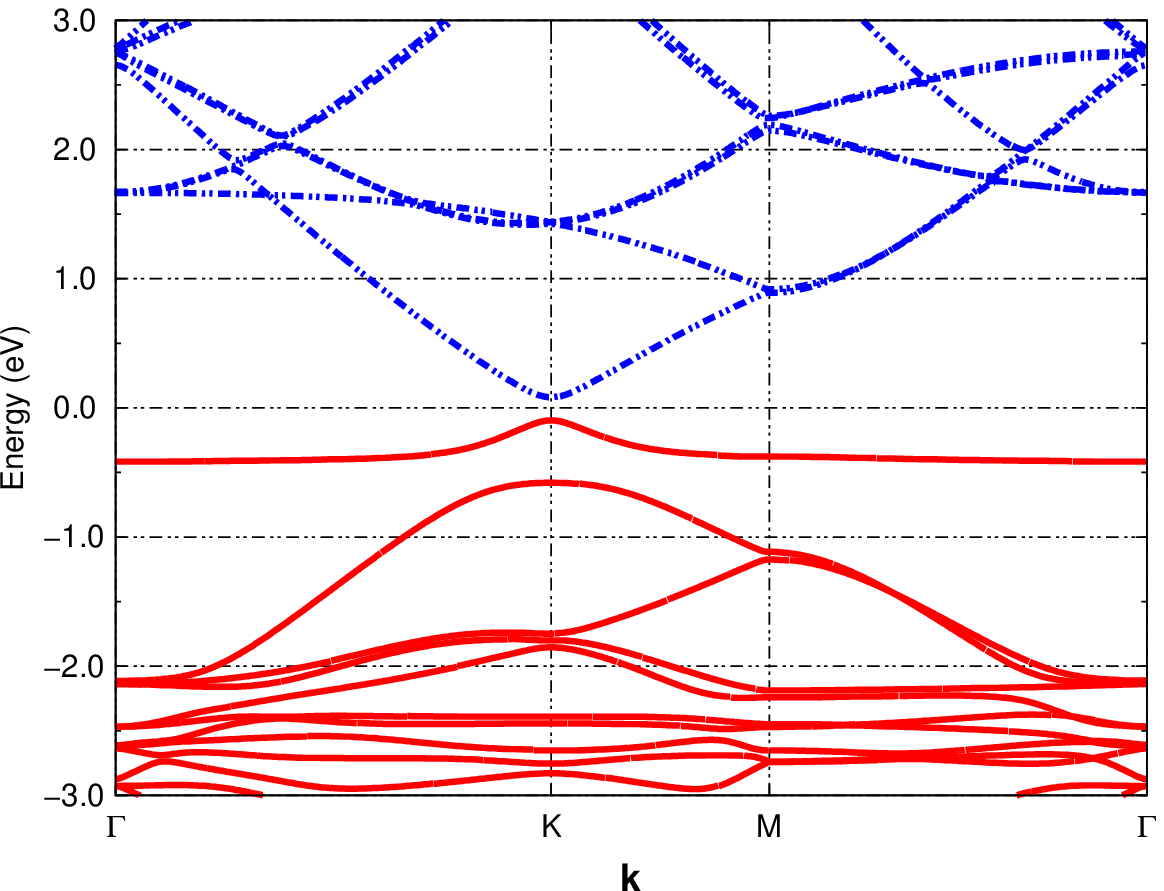}
b) \includegraphics[width=0.9\columnwidth,clip=true]{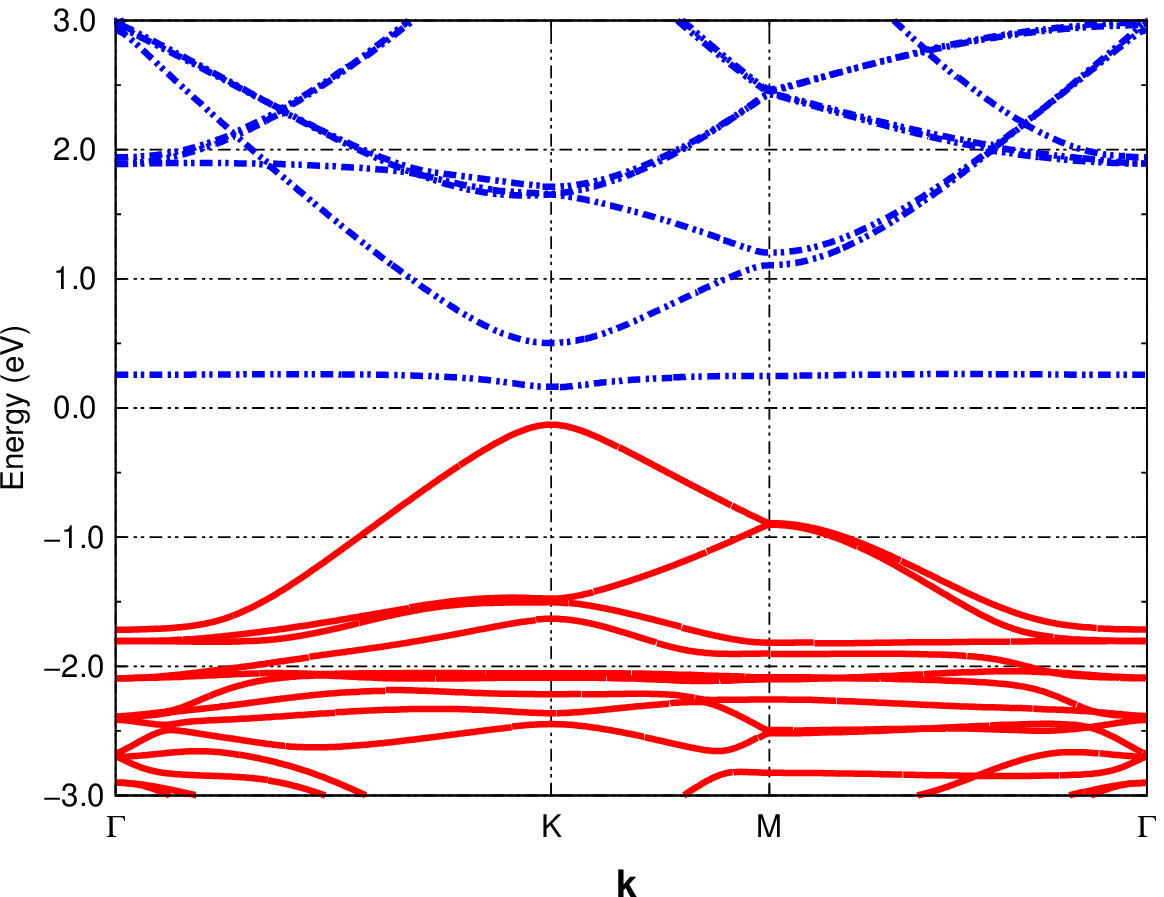}
\caption{\label{Au_bandst}(Colour online)
Spin-polarised band structures (eV) of Au in the vicinity of the Fermi energy.
(a) Majority spin band structure, and (b)  minority spin
band structure. The Fermi level is set to zero. Full lines denote occupied states
while dashed lines show empty levels.
%Au whose $6s$ level  crosses the graphene band structure below the Dirac point.
}
\end{center}
\end{figure}

We now turn to doping with Au.
Figure \ref{Au_bandst} a) and b) show the  majority and minority spin electronic band structures   for 
Au on graphene. Although there  are changes to the  dispersion of the bands   in the vicinity of the Dirac
point  there is no significant change
in the position of the Fermi level with respect to the Dirac point  compared with pristine graphene. This implies  that there is no 
significant  charge transfer between Au and graphene.

\begin{figure}
   \begin{center}
      \includegraphics[width=0.9\columnwidth,clip=true]{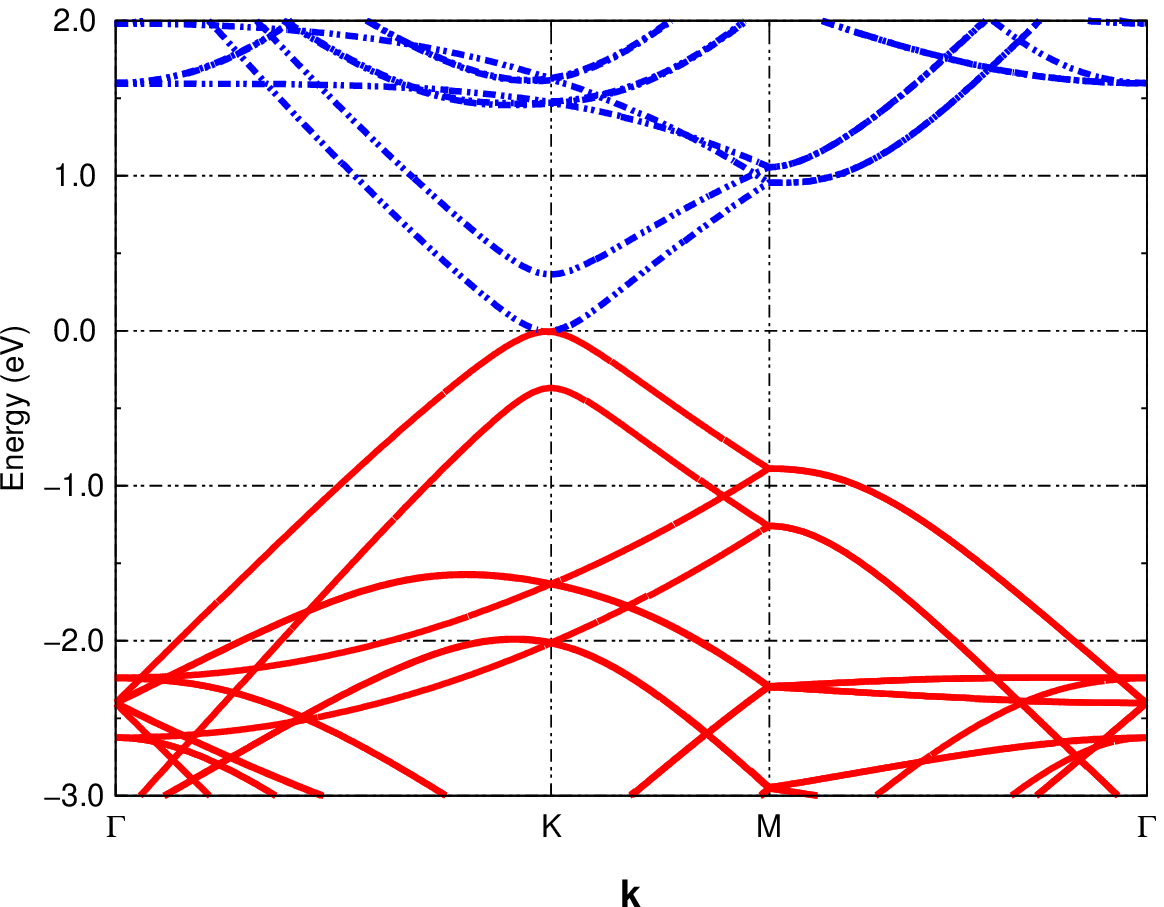}\\
      \includegraphics[width=0.9\columnwidth,clip=true]{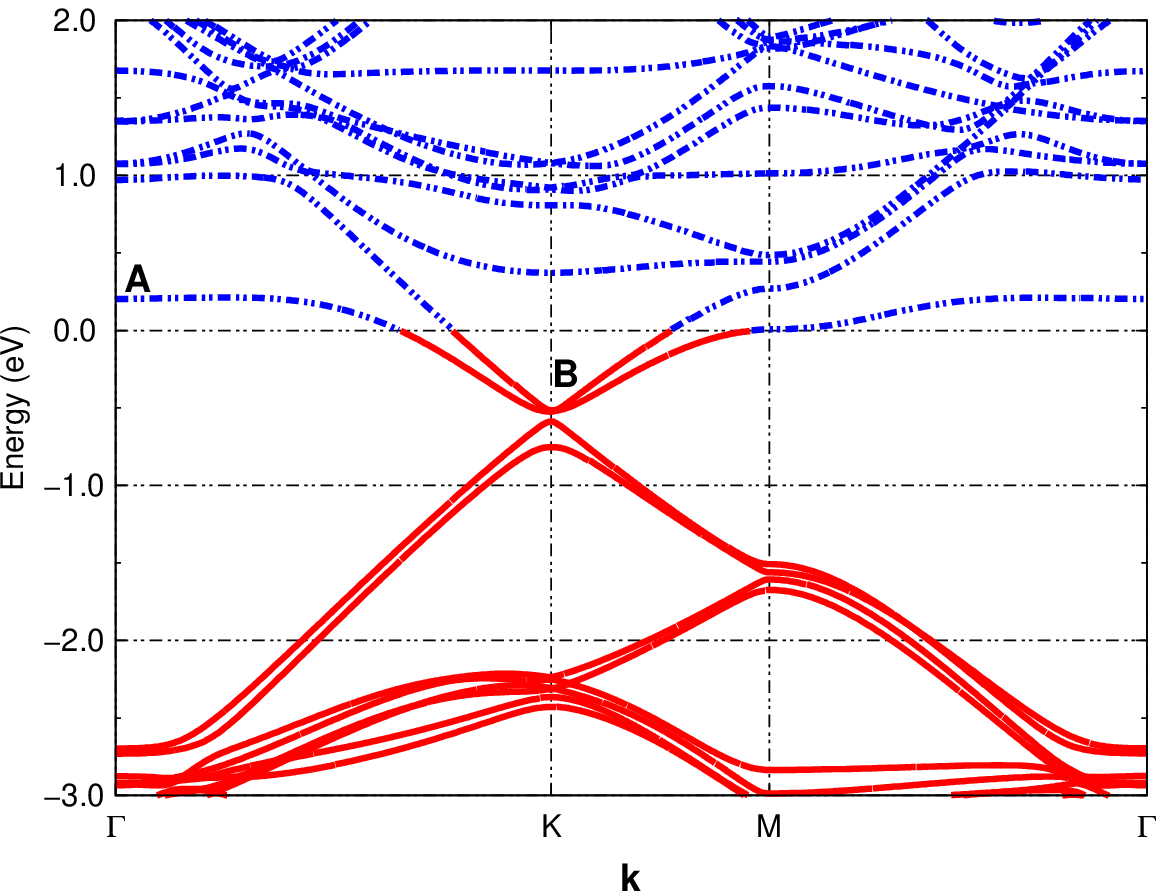}
      \caption{\label{Au_bilayer}(Colour online)
Electronic band structure (eV) for a  Au atom intercalated in  bilayer  graphene in the vicinity
of the Fermi energy. Full lines denote occupied states while dashed lines show empty levels.
The Fermi level is placed at zero. Note the band marked B  which is unoccupied in bilayer  graphene
is now  partially occupied showing doping has occurred.
}
   \end{center}
\end{figure}

To study the effect of additional graphene layers, a single Au atom was placed between
the AB stacked graphene bilayer. 
The relaxed ground state was spin-less and the Au atom   occupied a site nearly midway between the two graphene layers and 2.14 
\AA~directly above a   carbon atom in  the lower  layer and directly below the middle of the hexagon 
of the top layer. The separation between the graphene layers increased to approximately 4.5 \AA.
The graphene sheets do not appear to appreciably distort.
Figure \ref{Au_bilayer} a) and b) show the electronic band strucuture of a  pristine graphene bilayer and a
graphene bilayer intercalated with Au atoms.
As for   single layer graphene, the  highest occupied and lowest  empty  bands touch at the Dirac  point 
but the linear dispersion is lost.
However, when the Au atom is placed between the graphene layers, the Fermi level is shifted above the Dirac point. 
An analyis of the wavefunction shows the flat level far from the K point, marked as A,
results from the $6s$  Au level. When the  $6s$  level crosses the $\pi^*$ bands of graphene, they hybridise and 
charge is transferred to graphene effecting  n-type doping.
The level marked as B in Figure \ref{Au_bilayer}, is delocalised over graphene and this level is now occupied,
and unlike the case  when the second graphene sheet was absent.
We suggest that the effect of a second graphene layer 
is to compress the  Au $6s$- wavefunction with a consequent upward shift of the  level.
The same effect was observed when  ther second graphene layer was replaced by  a toluene molecule \cite{pinto}.

\section{Electrochemical doping}

Another   type of  doping involves  electrochemical redox 
reactions of graphene with water and adsorbates.
It has been reported that 
exposure of graphene  to a  humid  atmosphere   or  NO$_2$ 
 results in p-doping whereas exposure to gaseous 
NH$_3$ \cite{schedin-NM-6-652} and toluene results in n-doping
\cite{Kaverzin}. 
It is important  to realise  that other  carbon based materials 
can  exhibit similar effects. 
Diamond, whose surface has been treated with hydrogen
which reduces considerable its  work function, 
can also be doped p-type by exposure to a humid atmosphere \cite{maier-PRL-85-3472}. The effect is 
 suppressed by NH$_3$  and enhanced by NO$_2$ \cite{garrido}. Similar effects are seen in carbon nanotubes
\cite{kong-S-287-622,bradley-PRL-91-218301}.
These types of doping cannot be explained by 
an electronic mechanism involving   direct transfer of electrons from say toluene to graphene 
as the ionisation energy of toluene is $\sim 6$ eV  \cite{merkel} and greatly
exceeds the work function of graphene.
The  transfer  doping model  has similar difficulties in the case of NH$_3$. 
However, atmospheric doping has been  suggested as 
being  due to the  direct    transfer of 
charge  between graphene and defects  in  layers of ice on top of   the SiO$_2$ substrate \cite{wehling-APL-93-202110}.  
Nevertheless, this mechanism could not explain the 
surface doping  of diamond as 
SiO$_2$ is  not present.
Instead we argue here  that the  doping effects due to the atmosphere, NO$_2$ or
N$_2$O$_4$,
NH$_3$ and toluene  on   graphene that lead
to little change, or an increase, in  carrier  mobility  
are all examples of
electrochemical  redox reactions occurring in aqueous layers in contact
with graphene. A similar theory will also be applicable to 
nanotubes and already has been invoked for diamond \cite{chakrapani-s-30-1424}.
The  similar properties  of diamond films, nanotubes and graphene 
arise from their similar work functions although the large band
gap of diamond 
excludes the possibility of n-type doping.

The assumption that a water layer is in contact with 
graphene is a non-trivial one as 
theory has indicated that diamond and graphene are hydrophobic.
We are inclined to the view that  for graphene and nanotubes, likely places 
for water deposits are at the 
interfaces between  graphene  and SiO$_2$. 
Kelvin probe microscopy and X-ray spectroscopy have shown 
four or  five water layers exist on the surface of SiO$_2$ films grown  on Si
\cite{verdaguer-L-23-9699} and  there have been  FTIR observations   of
overtones of vibrations of water molecules
\cite{aarts-PRL-95-166104}
It is also widely believed that water can  exists in voids in  a-SiO$_2$  
\cite{bakos-PRB-69-195206}
and would be resistant to   thermal anneals if the voids are small. 
For diamond,  we suppose water is found at  places where the H termination has
been replaced by electrophilic  OH or oxygen groups.
However, further work is required on this point.

Redox reactions  are  reactions occurring in aqueous solution
involving changes in the charge states of the participants. 
The net charge change can be transferred to a electrode which in 
our case is graphene.
Whether such reactions occur require that the change in  the total Gibbs free energy  
is negative (described as a spontaneous reaction)  and that the  barriers are sufficiently small
that the reactions  can occur at room temperature. Tables of $\Delta G$ for a 
reaction in which unit  charge is removed to  or added from 
vacuum   are available  \cite{hand} but barrier heights are unknown and it must
be assumed that they are sufficiently low.  
The total Gibbs free energy change is then $\Delta G +W$  
for p-doping  or $\Delta G -W$ for n-doping. 
Here $W$ is the work function of graphene.

It has been argued that the  p-doping of diamond is due to
the electrochemical reduction of O$_2$  in the presence of water as in
\beq
O_2 + 2 H_2O +4 e = 4 OH^-
.
\eeq
The reaction 
is spontaneous   with the electron being 
transferred from the valence band of diamond and the formation of
 OH$^-$ groups \cite{chakrapani-s-30-1424}. A
 current carrying  hole is created near the diamond surface.
We now show  that an identical
reaction is spontaneous   
for graphene.
We consider the process occurs by removal of a electron 
from graphene to infinity followed by trapping of the
electron by O$_2$ and the subsequent  reaction with water. 
The extraction of the electron from graphene requires the expenditure
of  energy equal to its  work function $W$. This has been
measured for  graphite in air  to be 4.5 eV \cite{hansen-SS-481-172}.
This value is close to 4.9 eV  found for hydrogenated diamond \cite{rezek-APL-82-2266}
and we expect the same work function for graphene \cite{sque-PSSA-204-3078}.
The  similarlities of the work function of diamond, graphite and
nanotubes explains why  the same redox theory pertains to all
three materials.
%The redox reaction will be spontaneous if its 
%total free energy, $\Delta G  +W$ is negative.
Now $\Delta G$ is the free energy for the molecular reaction with the
charge brought  brought from  vacuum and can be written 
as $-eE-4.4$ eV  where $E$ is the  electrode potential
relative to a hydrogen electrode. The reaction is
assumed to involve molar concentrations of reactants and 
products. 
Standard electrode tables   give $E$ to be 
0.42 V   \cite{hand}. 
Thus the  total  free energy change
for a electron  to be taken from graphene 
and to reduce  the O$_2$/water couple  is  -0.42-4.4 +$W$  or -0.3 eV.
Thus the
reaction where electrons are extracted from graphene  by oxygen and water 
is spontaneous. 
%For H-terminated  diamond, there are conflicting values for the work function in
%the literature: The Munich group give 4.9 eV  but Chakrapani  gives  5.2 eV 
%\cite{chakrapani-s-30-1424}.
The above assumes that that    molar  concentration of OH$^-$  and O$_2$ 
are present 
but if this is not the case, then 
$\Delta G$  depends on the concentrations of OH$^-$ and O$_2$ with 
the reaction being inhibited in strongly basic conditions.
Nernst's equation gives $\Delta G$ to be 
\beq
\Delta G = -4.8 -0.059  (14-pH+ 0.25 \log_{10} (p_{O_2})) 
.
\eeq
Here $p_{O_2}$ is the partial pressure of oxygen in atm.
Taking this to be 0.21 atm gives   $\Delta G$ to vary from  -4.8 eV for 
$pH=14$ to -5.7 eV for $pH=1$. Acidic solutions will promote the reaction
as the concentration of OH$^-$ is so low. 
We note that OH$^-$ gradually builds up as doping increases
and might be expected to lead to reduced carrier mobility as explained above.
That this does not happen can be explained if OH$^-$
is either very effectively screened by surrounding SiO$_2$ groups or  is {\it 
mobile}
through SiO$_2$. Previous calculations
have shown that the latter is likely: the  binding of OH$^-$   to
the network is only 0.3 eV and thus  diffusion 
through SiO$_2$ could occur at room temperature \cite{bakos-PRB-69-195206}. 
Moreover, the theory  explains the hysteresis effect \cite{lohmann} if
the chemical reaction rates, or  diffusion  rate for  water, oxygen and OH$^-$, are slower than   the rate of 
change of  gate voltage. 
For positive gate voltages, the reaction product  OH$^-$  drifts away from 
graphene and the lower hydroxyl  concentration there encourages
further dissociation of water and hence  greater  increases in 
hole densities  in graphene
as indeed observed.
\cite{lohmann}.

The atmospheric  doping effect disappears  following vacuum annealing
around 200$^\circ$C  for both graphene 
and diamond \cite{maier-PRL-85-3472}.
We suppose this is because of the loss of oxygen 
and not to the loss of water. 
This is because we require a  water layer to explain 
doping  of annealed samples by gaseous  NO$_2$, NH$_3$ and toluene.

It might be thought that the effects of NO$_2$ and NH$_3$ can now
be explained using the same redox couple but taking into account the
changed $pH$. However this  need not be the case and  it is possible 
that different redox reactions occur especially in anaerobic conditions.
A possible  reaction involves the reduction of N$_2$O$_4$ as in 
$N_2O_4 +2e = 2 NO_2^-$ with a electrode potential of  $E = .87$ V
\cite{hand}. Thus $\Delta G $ would be -5.3 eV and  certainly 
could  account  for p-doping of both graphene and diamond. 

%However for basic conditions encountered for
%ammonia doping, taking $pH=14$ gives for the oxygen-water couple  $\Delta G  $ to be -4.8 eV and
%hence a reaction resulting in  {\it holes}  in graphene is spontaneous
%by 0.3 eV. However, what is  observed is n-doping in conflict with this 
%result. This is because the absence of oxygen (which has been
%annealed) now permits 
%an alternative couple  to dominate. 
%This involves the oxidation of NH$_3$ or NH$_4$OH. 

For the oxidation of   NH$_3$ or NH$_4$OH we have,
$2 NH_4 OH = N_2 + 2 H_2O +6 H^+ +  6 e $ with  $E =- 0.09 $ V and
yielding $\Delta G =4.5$ eV.
The electron produced by this reaction is transferred to graphene
and the change in the total  free energy is $\Delta G-W$. 
Thus the oxidation reaction leads to the supply of electrons
to  graphene  provided  that  the work function is  greater than 4.5 eV which is just possible.
However, we have ignored the basic conditions present  as the reaction is
inhibited in acidic conditions.
Nernst's equation gives $\Delta G $ ranging from 4.5 eV ($pH=1$) to
3.7 eV ($pH=14$). Thus the energy cost in removing an electron
from aqueous NH$_3$ is 3.7 eV and is easily recovered
by  its transfer to graphene.
The reaction occurs because of the scarcity of H$^+$ and 
a  very high N$_2$ bond energy.
It is noteworthy that this reaction is used industrially  to remove ammonia 
during waste water treatment.

%A better way of writing the oxidation reaction to exploit
%its connection with toluene  is
%$ NH_3 + 3 OH^-  = (1/2) N_2 + 3 H_2O +3 e $
%This is the sum of the 2 reactions
%\beq
%2 NH_4OH =N_2 + 2 H_2O +6 H^+  + 6 e  ,~  E =- 0.09 \\
%6 H^+ + 6 OH^- = 6 H_2O , E = 0.8 \\
%\eeq
%$\Delta G$ for first is 4.5 eV and for second is -0.8 eV.
%Hence for the reaction sum   $\Delta G = 3.7$, and the same as before.
%The  
%subsequent capture of   electrons by graphene gives a  net 
%$\Delta G + W = 3.7-4.5= - 0.8$ eV.
 
%Thus the oxidation  reaction
%\beq
%2NH_4OH + 6 OH^- =  N_2 + 8H_2O + 6 e  
%\eeq
%is spontaneous on graphene.

Finally we turn to toluene which has been found to result in n-type doping, a pronounced hysteresis effect and 
enhanced carrier mobility \cite{Kaverzin}.
One possibility is that this involves the redox couple where 
toluene is oxidised to benzyl alcohol.
Such a reaction has been reported previously \cite{tomat}.
From the tables of free energies \cite{hand},$\Delta G$ for 
 Toluene  $+ 2 OH^-  =  {\rm  Benzyl~alcohol} +{\rm {  H_2O}}  +2e$ is   3.95 eV.
%To obtain  the electrode  potential we consider the following 
%intermediate reactions.
%\beq
%H_2O_2 + 2 H^+ + 2e = 2H_2O, ~   E= 1.776 \\  
%2H_2O + 2 e = H_2 + 2 OH^-,  ~ E = -.83   \\
%2H^+ + 2 e = H_2, ~  E=0
%\eeq
%It is best to write these in terms of $\Delta G$ requiring a charge transfer of a single
%electron.
%\beq
%(1/2)H_2O_2 +  H^+ +  e = H_2O, ~  \Delta G =-6.176 \\     
%H_2O +  e =(1/2) H_2 +  OH^-,  ~ \Delta G =-3.57    \\
%H^+ +  e =  (1/2) H_2, ~   \Delta G  = -4.4 
%\eeq
%Adding the first two reactions gives
%$(1/2)H_2O_2 +   H^+ + 2e = (1/2) H_2 +  OH^-, ~ \Delta G = -9.746 $ eV
%and  subtracting  the third gives 
%$ (1/2) H_2O_2 + e =  OH^-, (A)~ \Delta G =  -5.35 $~ eV.
%Now $\Delta H$ for the reaction $H_2O_2 +$ Toluene = Benzyl alcohol + $H_2O$ is 
%-2.8 eV when carried out in liquids  and we assume the entropies balance
%on both sides, so $\Delta H = \Delta G$  and  thus
%$\Delta G$  for
%1/2 Toluene +  1/2 H$_2$O$_2$ = 1/2 Benzyl alcohol + 1/2  $H_2O$  (B)
%is -1.4 eV.
%Subtract A from B gives 
%$\Delta G$ for 
%$OH^- + 1/2 {\rm Toluene} =  1/2{\rm  BZ} +H_2O  +e$ is    3.95 eV.
Thus the electrochemical  oxidation of toluene to benzyl alcohol is 
spontaneous  as the work functions exceeds 3.95 eV and  
the liberated electron will be trapped by graphene achieving
n-type doping. For negative bias condition, OH$^-$ will  drift through
the SiO$_2$ substrate  towards graphene, removing a source of 
scattering centres   and leading to an increase in mobility

%The consumption of OH$^-$ would lead to an increase
%in the  hole mobility and the remaining 
 %OH$^-$
%n the SiO$_2$ would drift  to  the graphene under the 
%influence of the negative bias.

The  reaction depends  however on the concentration of OH$^-$ in the SiO$_2$ 
which is very low.
Another possibility  is the 	anaerobic
oxidation of toluene by water.
This reaction is known to occur naturally at room temperature   where it is catalysed by   microorganisms.
The standard electrode potential  for 
  Toluene +  21  H$_2$O $ \rightarrow $ 7 HCO$_3^-$ +43 H$^+  + 36 e$
is  0.27  V \cite{loveley}.  
In nature, the charge exchange takes place with the Fe$^{2+}$/Fe$^{3+}$ couple
\cite{loveley}. Taking into account  the standard electrode potential of
Fe$^{2+}$/Fe$^{3+}$ is 0.77 V, this  makes  the total free energy change for 
  Toluene +  21  H$_2$O  + 36 Fe$^{3+} \rightarrow $ 7 HCO$_3^-$ +43 H$^+$ 
+36 Fe$^{2+}$ 
to be -1.04 eV per electron. 
However, the reaction remains 
exothermic if the Fe$^{2+}$/Fe$^{3+}$ couple
is replaced by graphene,
as then  $\Delta G -W$  is -0.37 eV. 
It is worth emphasising that redox reactions are slow and 
thus graphene needs to be exposed to toluene for long periods
$\sim 1$ hour for a doping affect to be seen.
%$ {1 \over 36}{\rm{ Toluene +  Fe}}^{3+} +{21\over 36} {\rm H}_2O \rightarrow {7 \over 36} {\rm{HCO}}_3^- +{43\over 36} {\rm H}^+ 
%+ {\rm { Fe}}^{2+}$
Thus the anaerobic conversion of toluene into CO$_2$  in 
the presence of graphene is favoured.
To account for a mobility increase, we suppose that mobile   H$^+$ ions 
neutralise charged  scattering centres present  in the undoped graphene.

\section{Conclusions}

In summary, we conclude there are two types of dopants for graphene.
The first which can be called {\it electronic doping} 
occurs when there is a direct exchange of electrons with an adsorbate 
and graphene. A good example is K.
Such doping occurs   promptly and leads to  a reduction in carrier 
mobility and there should be no hysteresis effects.
The second, involving {\it electrochemical doping} 
occurs by redox reactions involving water at the interface. This 
can lead to   an increase in carrier mobility but  
requires  appreciable times to occur. This leads to  hysterisis effects.
The assumption that  one  or  both  of the  charged products OH$^-$ or H$^+$ is
mobile
and responds to the field due to the gate voltage  could explain  increases in   carrier mobility.

\section*{Acknowledgements}
The authors are thankful to Derek Palmer for the helpful discussions.

\end{document}